\documentclass[sigconf]{acmart}

\usepackage[utf8]{inputenc} %
\usepackage[T1]{fontenc}    %
\usepackage{hyperref}       %

\usepackage{hyperxmp}
\usepackage{url}            %
\usepackage{booktabs}       %
\usepackage{amsfonts}       %
\usepackage{nicefrac}       %
\usepackage{microtype}      %
\usepackage{xcolor}         %

\usepackage{multirow}
\usepackage{subfigure}

\usepackage[english]{babel}
\usepackage{moresize}
\usepackage{amsmath}
\usepackage{algorithmic}
\usepackage{balance}
\usepackage{comment}
\usepackage{paralist}
\usepackage{bm}
\usepackage{pgfplots}
\usetikzlibrary{pgfplots.dateplot}

\usepackage{flushend}
\usepackage[english]{babel}
\usepackage{graphicx}

\usepackage{amssymb}
\usepackage{amsfonts}
\usepackage{url}
\usepackage{bbm}
\usepackage{longtable}
\usepackage{rotating}
\usepackage{multirow}
\usepackage{mathrsfs}
\usepackage{enumitem}
\usepackage[linesnumbered,algoruled,boxed,lined]{algorithm2e}
\usepackage{adjustbox}
\usepackage{hyperref}
\usepackage{pgfplots}
\usetikzlibrary{pgfplots.dateplot}
\usepackage{filecontents}

\newcommand{\ie}{\textit{i}.\textit{e}.}
\newcommand{\eg}{\textit{e}.\textit{g}.} 
\newcommand{\wrt}{\textit{w}.\textit{r}.\textit{t}}

\newcommand{\hide}[1]{} %

\newcommand{\ve}{\textbf{e}}
\newcommand{\vece}{\textbf{e}}
\newcommand{\vecv}{\textbf{v}}
\newcommand{\setu}{\mathcal{U}}
\newcommand{\setv}{\mathcal{V}}
\newcommand{\sete}{\mathcal{E}}
\newcommand{\setg}{\mathcal{G}}
\newcommand{\sets}{\mathcal{S}}

\newcommand{\mata}{\textbf{A}}
\newcommand{\matd}{\textbf{D}}
\newcommand{\mate}{\textbf{E}}

\newcommand{\mats}{\textbf{S}}
\newcommand{\matw}{\textbf{W}}
\newcommand{\domr}{\mathbb{R}}

\def\model{LightGNN}

\AtBeginDocument{%
  }

\copyrightyear{2025}
\acmYear{2025}
\setcopyright{acmlicensed}\acmConference[WSDM '25]{Proceedings of the Eighteenth ACM International Conference on Web Search and Data Mining}{March 10--14, 2025}{Hannover, Germany}
\acmBooktitle{Proceedings of the Eighteenth ACM International Conference on Web Search and Data Mining (WSDM '25), March 10--14, 2025, Hannover, Germany}
\acmDOI{10.1145/3701551.3703536}
\acmISBN{979-8-4007-1329-3/25/03}

\begin{document}



\title{LightGNN: Simple Graph Neural Network for Recommendation}

\author{Guoxuan Chen}
\affiliation{%
  \institution{University of Hong Kong}
  \city{Hong Kong}
  \country{China}
  }
\email{guoxchen@foxmail.com}

\author{Lianghao Xia}
\affiliation{%
  \institution{University of Hong Kong}
  \city{Hong Kong}
  \country{China}
  }
\email{aka\_xia@foxmail.com}

\author{Chao Huang}
\authornote{Chao Huang is the Corresponding Author.}
\affiliation{%
  \institution{University of Hong Kong}
  \city{Hong Kong}
  \country{China}
  }
\email{chaohuang75@gmail.com}

\renewcommand{\shortauthors}{Guoxuan Chen, Lianghao Xia, and Chao Huang}

\begin{abstract}
Graph neural networks (GNNs) have demonstrated superior performance in collaborative recommendation through their ability to conduct high-order representation smoothing, effectively capturing structural information within users' interaction patterns. However, existing GNN paradigms face significant challenges in scalability and robustness when handling large-scale, noisy, and real-world datasets. To address these challenges, we present \model, a lightweight and distillation-based GNN pruning framework designed to substantially reduce model complexity while preserving essential collaboration modeling capabilities. Our \model\ framework introduces a computationally efficient pruning module that adaptively identifies and removes redundant edges and embedding entries for model compression. The framework is guided by a resource-friendly hierarchical knowledge distillation objective, whose intermediate layer augments the observed graph to maintain performance, particularly in high-rate compression scenarios. Extensive experiments on public datasets demonstrate \model's effectiveness, significantly improving both computational efficiency and recommendation accuracy. Notably, \model\ achieves an 80\% reduction in edge count and 90\% reduction in embedding entries while maintaining performance comparable to more complex state-of-the-art baselines. The implementation of our \model\ framework is available at the github repository: \textcolor{blue}{\url{https://github.com/HKUDS/LightGNN}}.
\end{abstract}

\maketitle

\section{Introduction}
\label{sec:intro}
Recommender systems~\cite{zhang2019deep, gao2023survey} have become indispensable in modern online platforms, effectively addressing information overload and enhancing user engagement through personalized service delivery. At the core of these systems, Collaborative Filtering (CF)~\cite{su2009survey, koren2021advances} stands as a dominant paradigm, leveraging users' historical interactions to model latent preferences for behavior prediction.

The evolution of collaborative filtering has spawned diverse approaches, from classical matrix factorization methods (\eg~\cite{koren2009matrix}) to sophisticated neural architectures (\eg~\cite{he2017neural}). Among these developments, Graph Neural Networks (GNNs) have emerged as particularly powerful tools for CF-based recommendation, distinguished by their ability to capture complex, high-order interaction patterns through iterative embedding smoothing. Pioneering works include NGCF~\cite{wang2019neural}, which introduced graph convolutional networks (GCNs) to model user-item relationships, and LightGCN~\cite{he2020lightgcn}, which simplifies GCNs to their essential components for recommendation. To address the challenge of sparse interactions in GNN-based recommendation, researchers have developed innovative self-supervised learning (SSL) techniques, including SGL~\cite{wu2021self}, NCL~\cite{lin2022improving}, and HCCF~\cite{xia2022hypergraph}. These approaches significantly enhance recommendation accuracy by leveraging self-augmented supervision signals.

Despite significant advancements in GNNs, we would like to emphasize two inherent limitations that continue to challenge GNN-based CF models. \textbf{i) Limited scalability of GNNs}: Online recommendation services typically handle vast amounts of relational data (e.g., millions of interactions). This causes the size of user-item graphs to increase dramatically, resulting in a considerable number of information propagation operations within GNNs. Such scalability issues present challenges concerning storage, computational time, and memory requirements. Furthermore, GNN-based CF relies heavily on id-corresponding embeddings for user and item representation~\cite{he2020lightgcn}, with the complexity of these embeddings directly linked to the growing number of users and items, incurring significant memory costs. \textbf{ii) Presence of pervasive noise in interaction graphs}: Collaborative recommenders mainly utilize users' implicit feedback, such as clicks and purchases, because of its abundance. However, these interaction records often contain substantial noise that diverges from users' true preferences, including misclicks and popularity biases~\cite{wang2021denoising}. Although some existing methods address scalability through techniques like random dropping (e.g., PinSage~\cite{ying2018graph}) or knowledge distillation (KD) (e.g., SimRec~\cite{xia2023graph_less}), they remain susceptible to misinformation, which can result in inaccurate predictions from their compressed recommenders.

\begin{figure}[t]
    \centering
    \includegraphics[width=\columnwidth]{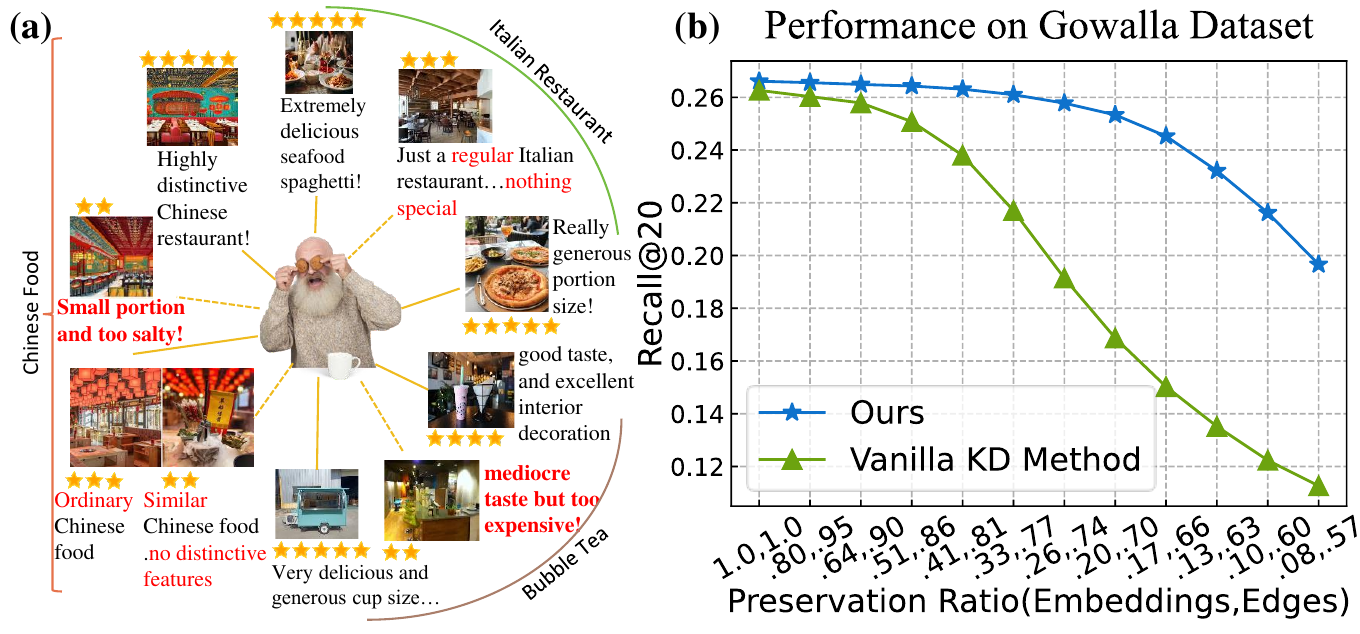}
    \vspace{-0.15in}
    \caption{Illustrations depicting (a) redundant and noisy user interactions, with red text indicating noisy feedback, and (b) the superior performance retention of \model\ compared to vanilla KD, especially under high-rate pruning.}
    \label{fig:intro}
    \vspace{-0.1in}
\end{figure}

To address these limitations, this paper proposes pruning redundant and noisy components in GNNs, specifically targeting graph edges and embedding entries. We aim to enhance model scalability while preserving essential user preference features. However, achieving this objective presents non-trivial challenges, outlined as:
\begin{itemize}[leftmargin=*]
    \item How to identify the graph edges and embedding entries that are genuinely redundant or noisy in the user-item interaction graph?\\\vspace{-0.12in}
    \item How to maintain the high performance of GNN-based CF when significant structural and node-specific information is removed?
\end{itemize}
As illustrated in Figure~\ref{fig:intro}(a), a considerable proportion of items that users interact with fall into the same category, leading to redundant information about users' preferences. By identifying and removing this redundancy from both structures and parameters, we can significantly reduce the complexity of GNN-based CF. Additionally, many observed interactions represent noise linked to users' negative feedback, as revealed by the review text. This noise can disrupt the preference modeling of existing compressed CF methods, which often fail to explicitly identify such noisy information. Regarding the second challenge, depicted in Figure~\ref{fig:intro}(b), traditional knowledge distillation approaches struggle to effectively maintain performance when compressing the GNN model at a high ratio due to the limited number of edges and parameters. In contrast, our innovative hierarchical KD offers enhanced preservation capabilities.

Fully aware of these challenges, we introduce a GNN pruning framework called \model\ that facilitates efficient and denoised recommendations. \model\ incorporates graph structure learning to explicitly assess the likelihood of redundancy or noise for each edge and embedding entry. This learning process is supervised in an end-to-end fashion, leveraging the downstream recommendation task alongside a hierarchical knowledge distillation paradigm. Inspired by the advantages of global relation learning in recommendation~\cite{xia2022hypergraph}, our KD approach features an intermediate distillation layer that utilizes high-order relations to enhance candidate edges in the compressed model. This augmentation improves the model's capacity to maintain recommendation performance under high-rate compression. Through innovative importance distillation and prediction-level and embedding-level alignments, our hierarchical knowledge distillation enriches learnable pruning with abundant supervisory signals, boosting its compression capability.

The contributions of our \model\ are summarized as follows:
\begin{itemize}[leftmargin=*]
    \item We introduce a novel GNN pruning framework for recommendation, explicitly identifying and eliminating redundancy and noise in GNNs to enable efficient and denoised recommendations.\\\vspace{-0.12in}
    \item Our \model\ framework integrates an innovative hierarchical knowledge distillation paradigm, seamlessly compressing GNNs at high ratios while preserving prediction accuracy.\\\vspace{-0.12in}
    \item We conduct extensive experiments to demonstrate the superiority of \model\ in terms of recommendation accuracy, inference efficiency, model robustness, and interpretability.
\end{itemize}

\section{GNN-based Collaborative Filtering}
\label{sec:model}

Graph neural network (GNN) has been shown a most effective solution to collaborative filtering (CF)~\cite{chen2020revisiting, xia2023automated}. The CF task typically involves a user set $\mathcal{U}$ ($|\mathcal{U}|=I$), an item set $\setv$ ($|\setv|=J$), and a user-item interaction matrix $\textbf{A}\in\mathbb{R}^{I\times J}$. For a user $u_i\in\mathcal{U}$ and an item $v_j\in\setv$, the entry $a_{i,j}\in\textbf{A}$ equals $1$ if user $u_i$ has interacted with item $v_j$, otherwise $a_{i,j}=0$. Common interactions include users' rating, views, and purchases. GNN-based CF methods construct the user-item graph based on the interaction matrix $\textbf{A}$. This graph can be denoted by $\mathcal{G}=(\mathcal{U}, \setv, \mathcal{E})$, where $\mathcal{U}, \setv$ serve as the graph vertices, and $\mathcal{E}$ denotes the edge set. For each $(u_i, v_j)$ that satisfies $a_{i,j}=1$, there exists bidirectional edges $(u_i, v_j), (v_j, u_i)\in\mathcal{E}$.

Based on the user-item graph $\mathcal{G}$, GNNs conduct information propagation to smooth user/item embeddings for better reflecting the interaction data. Specifically, it firstly assigns initial embeddings $\textbf{e}_i, \textbf{e}_j \in\mathbb{R}^d$ to each user $u_i$ and item $v_j$, respectively. Here $d$ represents the hidden dimensionality. Then it iteratively propagates each node's embedding to its neighboring nodes for representation smoothing. Take the widely applied LightGCN~\cite{he2020lightgcn} as an example, the embeddings for user $u_i$ and item $v_j$ in the $l$-th iteration are:
\begin{align}
    \textbf{e}_{i,l} = \sum_{(v_j, u_i)\in\mathcal{E}} \frac{1}{\sqrt{d_i d_j}}\textbf{e}_{j,l-1},~~~~~
    \textbf{e}_{j,l} = \sum_{(u_i, v_j)\in\mathcal{E}} \frac{1}{\sqrt{d_i d_j}}\textbf{e}_{i,l-1}
\end{align}
where $\textbf{e}_{i,l}, \textbf{e}_{i,l-1}\in\mathbb{R}^d$ denote the embedding vectors for $u_i$ in the $l$-th and the $(l-1)$-th iterations, and analogous notations are used in $\ve_{j,l}, \ve_{j,l-1}$. The $0$-th embedding vectors $\ve_{i,0}, \ve_{j,0}$ uses the initial embeddings $\ve_i, \ve_j$. And $d_i, d_j$ represent the degrees of nodes $u_i, v_j$, for Lapalacian normalization. After a total $L$ iterations, GNN-based CF aggregates the multi-order embeddings for final representations $\bar{\ve}_i, \bar{\ve}_j\in\mathbb{R}^d$ and user-item relation predictions $\hat{y}_{i,j}$, as follows:
\begin{align}
    \hat{y}_{i,j}=\bar{\ve}_i^\top \bar{\ve}_j, ~~~~\bar{\ve}_i=\sum_{l=0}^L \ve_{i,l}, ~~~~\bar{\ve}_j=\sum_{l=0}^L \ve_{j,l}
\end{align}
With the prediction scores $\hat{y}_{i,j}$, the GNN models are optimized by minimizing the BPR loss function~\cite{rendle2012bpr} over all positive user-item pairs $(u_i, v_{j^+})\in\mathcal{E}$, and sampled negative pairs $(u_i, v_{j^-})$, as follows:
\begin{align}
    \mathcal{L}_{bpr}=\sum_{(u_i,v_{j^+}, v_{j^-})} -\log \text{sigm} (\hat{y}_{i,j^+} - \hat{y}_{i,j^-})
\end{align}
Though the above GNN framework achieves state-of-the-art performance in recommendation, its scalability is limited by the large-scale interaction graph and embedding table. In light of this, this paper proposes \model\ aiming to effectively prune the GNN model for efficient graph neural collaborative filtering.

\section{Methodology}
\label{sec:solution}
\begin{figure*}[t]
    \begin{center}
    \includegraphics[width=0.99\linewidth]{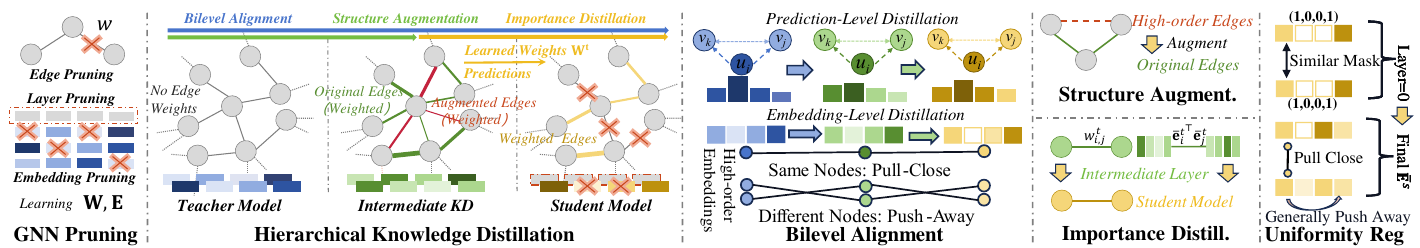}
    \end{center}
    \vspace{-0.1in}
    \caption{Overall framework of the proposed \model\ model.}
    \label{fig:framework}
    \vspace{-0.1in}
\end{figure*}

This section goes through the proposed \model\ to show the technical details. The overall framework is illustrated in Figure~\ref{fig:framework}.

\subsection{Graph Neural Network Pruning}
Inspired by the lottery ticket hypothesis for GNNs~\cite{frankle2018lottery, chen2021unified}, we propose to use only a subset of GNN's parameters that maximally preserve the model functionality, to improve its efficiency. Specifically, the time complexity for a typical GNN model as aforementioned is $\mathcal{O}(L\times |\sete|\times d)$, and the space complexity is correspondingly $\mathcal{O}(|\sete|+(I+J)\times d)$. Therefore by reducing the number of edges $|\sete|$, and the number of non-zero elements in the $d$ embedding dimensions, our \model\ is able to optimize both the computational efficiency and memory efficiency. To achieve this, it is essential to identify the noisy and redundant parts in the edges $\sete$ and the embedding table $\textbf{E}=\{\ve_i,\ve_j|u_i\in\setu, v_j\in\setv\}$, to prevent performance degradation.

\subsubsection{\bf Edge Pruning}
To this end, \model\ employs a sparse weight matrix $\matw\in\domr^{I\times J}$ for edge pruning. If an edge $(u_i, v_j)$ is a candidate for pruning, the corresponding weight $w_{i,j}$ in $\matw$ is a learnable parameter. Otherwise $w_{i,j}$ is set as 0 and is not optimized. With the weight matrix $\matw$, the graph information propagation process for the pruned GNN is conducted as follows:
\begin{align}
    \label{eq:prune_forward}
    \mate_{\setu, l} = \matd_\setu^{-\frac{1}{2}} \cdot (\mata \odot \matw)\cdot \matd_\setv^{-\frac{1}{2}} \cdot \mate_{\setv, l-1} + \mate_{\setu, l-1}
\end{align}
where $\odot$ denotes the element-wise product operator which injects the learnable weights $\matw$ into the information propagation process. Here $\mate_{\setu,l}, \mate_{\setu,l-1}\in\domr^{I\times d}$ denote the user embedding table in the $l$-th and the $(l-1)$-th iteration, and $\mate_{\setv, l-1}\in\domr^{J\times d}$ denotes the embedding matrix for items in the $(l-1)$-th iteration. And $\matd_\setu\in\domr^{I\times I}, \matd_\setv\in\domr^{J\times J}$ denote the degree matrices for users and items, respectively. The information propagation to obtain higher-order item embeddings $\mate_{\setv,l}$ is analogously using $(\mata\odot\matw)^\top$.

Based on the parametric information propagation, the weights $\matw$ participate in the calculation for final user/item embeddings, which are then used for predictions and loss calculations. Through the back propagation, $\matw$ is tuned to reflect the importance of edges, wherein larger $|w_{i,j}|$ denotes the edge $(u_i, v_j)$ having a larger influence on producing better recommendation results. 
In light of this property, our \model\ framework prunes the less important edges (noises or redundancies) after training, specifically by setting the $\rho$\% candidate edges with the least importance to 0 (see~\ref{sec:imp_kd}), where $\rho\in(0,100)$ denotes the proportion to drop. The pruning algorithm follows an iterative manner with multiple runs. In each run, \model\ first conducts parameter optimization for model training and pruning weight tuning, and then prunes the GNN by dropping edges and other parameters.

\subsubsection{\bf Embedding and Layer Pruning}
As indicated by the complexity analysis for GNNs, the parameters for representing users and items (\ie~embeddings $\mate$) also contribute significantly to the running time and the memory costs of GNNs. Therefore \model\ follows the similar pruning algorithm for edges to prune the entries in the embedding matrix $\mate$. As the scalar parameters in $\mate$ already reflect the importance of their corresponding entries, \model\ does not employ extra pruning weights for embeddings. Analogously, \model\ alternately conducts model training and parameter pruning with ratio $\rho'$\% according to the absolute value $|e_{i,d'}|$, where $e_{i,d'}$ represents the $d'$-th dimension in $i$'s embedding vector.

In addition to the edges and embeddings, the time complexity of GNNs suggests that the number of graph propagation layers $L$ also greatly impacts the computation time of GNNs. Moreover, in practice, $L$ is also significant to influence the temporary memory costs for stacking the intermediate results. Thus our \model\ further reduces the number of graph iterations $L$ for efficiency, which also alleviates the over-smoothing effect of GNNs~\cite{xia2022hypergraph}.

\subsection{Hierarchical Knowledge Distillation}
\subsubsection{\bf Bilevel Alignment}
Motivated by the strength of knowledge distillation (KD) in compressing the learned knowledge of advanced models into light-weight architectures~\cite{xia2023graph_less}, the proposed \model\ develops a hierarchical knowledge distillation framework to maximally retain the original high performance in the pruned GNN model.
Taking a well-trained GNN model (\eg~LightGCN~\cite{he2020lightgcn}) as the teacher, \model\ aligns the student model with pruned structures, embeddings, and GNN layers to the teacher model with respect to both hidden embeddings and final predictions. In the prediction level, the following loss function is applied:
\begin{align}
    &\mathcal{L}_{p-kd}=\sum_{\vecv} - \Bigl(\sigma(\epsilon^t_{\vecv} / \tau) \cdot \log \sigma(\epsilon^s_{\vecv} / \tau)
    + \overline{\sigma}(\epsilon^t_{\vecv} / \tau)) \cdot \log \overline{\sigma}(\epsilon^s_{\vecv} / \tau) \Bigl)\nonumber\\
    &\text{where}~~\vecv=(u_i,v_{j^1}, v_{j^2}),~~\overline{\sigma}(x) = 1 - \sigma(x), ~~ \epsilon_\vecv^* = \hat{y}_{i,j^1}^* - \hat{y}^*_{i,j^2}
\end{align}
Here $(u_i, v_{j^1}, v_{j^2})$ denotes the randomly sampled training tuples analogous to the BPR loss, while $v_{j^1}$ and $v_{j^2}$ are not fixed to be positive or negative samples. $\sigma(\cdot)$ denotes the sigmoid function to constrain the values to be within $(0, 1)$. And $\tau\in\domr$ is known as the temperature coefficient~\cite{hinton2015distilling}. We denote the predictions made by the student model using the superscript $s$, and denote the predictions made by the teacher model with the superscript $t$. With this training objective, our \model\ framework encourages the pruned GNN model to mimic the predictions made by the complete GNN model with all the edges, embedding entries and propagation iterations, to obtain the teacher's prediction ability as much as possible.

Besides the prediction-level alignment, our \model\ aligns the teacher model and the student model by treating their learned embeddings as paired data views for contrastive learning. In specific, the following infoNCE loss function~\cite{oord2018representation} is applied:
\begin{align}
    \mathcal{L}_{e-kd}=&-\sum_{u_i\in\setu}  \log \text{softmax}(\mats_\setu, u_i) - \sum_{v_j\in\setv}  \log \text{softmax}(\mats_\setv, v_j)\nonumber\\
    \text{where}~~&\text{softmax}(\mats_\setu, u_i) = \frac{\exp s_{i,i}}{\sum_{u_{i'}} \exp s_{i',i}},~~
    s_{i',i} = \cos(\bar{\vece}_{i'}^{s}, \bar{\vece}_{i}^t)%
\end{align}
Here $s_{i',i}\in\mats_\setu$ denotes the cosine similarity between the final embeddings $\bar\vece_{i'}^s, \bar\vece_{i}^t$ for the users $u_{i'}$ and $u_{i}$, given by the student model and the teacher model, respectively. The item-side embedding-level KD is calculated analogously. With this embedding-level KD objective, our \model\ can better guide the pruned GNN to preserve the essential graph structures and parameters in a deeper level.

\subsubsection{\bf Intermediate KD Layer for Structure Augmentation}
Due to the sparsity nature of the user-item interaction data, some key preference patterns are not reflected by the direct neighboring relations but preserved by the high-order relations. To facilitate the capturing of these high-order connections during our edge pruning, we augment the knowledge distillation of \model\ with an intermediate KD layer model for edge augmentation.

To be specific, \model\ conducts a two-stage distillation, firstly from the original GNN to an augmented GNN, and then from the augmented GNN to the final pruned GNN. The augmented GNN does not prune any edges or embedding entries, but instead includes the high-order connections as augmented edges. Formally, the augmented GNN has the same model architecture (Eq. \ref{eq:prune_forward}) as the student but works over the following augmented interaction graph:
\begin{align}
    \bar{\setg} = (\setu, \setv, \bar{\sete}),~~~~~\bar{\sete}=\{(u_i,v_j),(v_j, u_i)|\bar{a}_{i,j}^{(h)}	\neq 0 \}
\end{align}
where $\bar{a}_{i,j}^{(h)}$ denotes the entry for $(u_i, v_j)$ in the $h$-th power of the symmetric adjacent matrix with self loop~\cite{wang2019neural}. In other words, edge $(u_i, v_j)$ exists in the augmented graph $\bar{\setg}$ if $u_i$ can be connected to $v_j$ via any path with its length shorter than or equal to $h$ hops in the original graph. With this structure augmentation, the augmented GNN directly includes the high-order connections in the model parameters, to prevent losing the key high-order patterns in radical edge pruning. During the intermediate KD, the augmented GNN is supervised by the original GNN (no weights), not only to mimic its accurate predictions, but also to learn proper weights $\matw^{t}$ for all the edges. The intermediate KD layer prevents the augmented larger graph from introducing noises using the supervision of the bilevel distillation from original GNN and the adaptive edge weights.
\subsubsection{\bf Importance Distillation for Pruning}
\label{sec:imp_kd}
After the first knowledge distillation from the original GNN to the augmented GNN model, our \model\ then distills its learned knowledge with structure augmentation to the final pruned GNN model. Apart from the aforementioned bilevel alignment, \model\ further enhances this second KD with the importance distillation, which explicitly leverages the learned importance weights in the intermediate model to increase the precision of pruning weights in the final model. Specifically, the pruning weight matrix in the final pruned GNN is a compound variable whose entries are calculated as follows:
\begin{align}
    \bar{w}_{i,j}^s = w_{i,j}^s + \beta_1 \cdot w_{i,j}^t + \beta_2 \cdot \sigma(\bar{\vece}^{t\top}_i \bar{\vece}_j^t)~~~ \text{for}~(u_i, v_j)\in\sete
    \label{eq:imp_distill}
\end{align}
where $\bar{w}_{i,j}^s\in\domr$ denotes the weight to decide if edge $(u_i, v_j)$ should be pruned, and it is acquired using the independent edge weight $w_{i,j}^s\in\matw^s$ of the final student model, the tuned edge weight $w_{i,j}^t\in\matw^t$ of the intermediate GNN as the teacher model, and the edge prediction made by the intermediate GNN's final embeddings $\bar{\vece}_i^t, \bar{\vece}_j^t\in\domr^d$. Here $\beta_1, \beta_2$ denote two hyperparameters for weighting and we define the sparse decision matrix $\bar{\matw}^s=\{\bar{w}_{i,j}^s\}_{I \times J}$.

With this importance distillation in the edge pruning, the pruning weights $\bar{\matw}^s$ in the final student model are not only trained in the end-to-end manner using the bilevel KD objectives, but also directly adjusted by the well-trained weights in the intermediate teacher model. Moreover, by utilizing the edge weights obtained in the augmented graph, the pruned GNN is injected with the high-order connectivity to facilitate edge dropping and global relation learning.
It is worth noting that, apart from the edge pruning, the student's edge weights are also employed in the graph information propagation, to enrich the pruned GNN with less edges but compensatory, adaptive and informative edge importance.

\subsection{Optimization with Uniformity Constraint}
Inspired by the advantage of learning uniform embeddings in CF~\cite{xia2023automated, wang2022towards}, our \model\ proposes to regularize the model optimization with an adaptive uniformity constraint based on contrastive learning.
In specific, the constraint minimizes the pairwise inner-product between embeddings to enforce representation uniformity, while maximizing the embedding similarity between nodes with similar pruning masks. In this way, the positive relations are augmented by the learned pruning weights for enhancement. Formally, the adaptive uniformity constraint is as follows:
\begin{align}
    \mathcal{L}_{u-reg} = &\sum_{u_i\in\mathcal{U}} \left(-\log
    \frac{ \sum_{u_{i^1}\in\sets_{i}}~ \exp\left( {\bar{\vece}^s_i}{}^{\top} {\bar{\vece}}^s_{i^1} / \tau \right)}{ \sum_{u_{i^2}\in\setu} ~\exp \left( {\bar{\vece}^s_i}{}^\top \bar{\vece}^s_{i^2} /\tau\right) }\right)\nonumber\\
    +&\sum_{v_j\in\setv} \left(-\log
    \frac{ \sum_{v_{j^1}\in\sets_j} ~\exp\left( {\bar{\vece}^s_j}{}^\top \bar{\vece}^s_{j^1} / \tau \right)}{ \sum_{v_{j^2}\in\setv} ~\exp \left({\bar{\vece}^s_j}{}^\top \bar{\vece}^s_{j^2} /\tau \right)}\right)
\end{align}
where $\sets_i$ and $\sets_j$ denote the positive sets of user $u_i$ and item $v_j$, respectively, which are determined by picking the users/items that share the highest similarity in embedding pruning. Take the user side as an example, the neighborhood set $\sets_i$ is acquired by:
\begin{align}
    \label{eq:pos_samples}
    \sets_i=\left\{u_{i^1} \: \big| \: \|\mathbbm{e}_i\odot \mathbbm{e}_{i^1}\|_0 \geq \max\left( \|\mathbbm{e}_i\|_0, \|\mathbbm{e}_{i^1}\|_0 \right) - \delta  \right\}
\end{align}
where $\mathbbm{e}_i, \mathbbm{e}_{i^1}\in\{0,1\}^d$ denote binary pruning masks for the $0$-th embedding vectors $\vece_i^s$ and $\vece_{i^1}^s$, respectively.
Operator $\odot$ denotes the element-wise multiplication, and $\|*\|_0$ denotes the $l_0$ norm of vectors. $\delta$ represents the threshold hyperparameter for similarity relaxation, which is selected according to the pruning ratio.

With the above contrastive loss using similarly-pruned embeddings as positive sets, \model\ can learn uniformly-distributed embeddings while capturing the node-wise similarity during the pruning process. Combining it with the collaborative filtering loss $\mathcal{L}_{bpr}$, the bilevel KD losses $\mathcal{L}_{p-kd}$ and $\mathcal{L}_{e-kd}$, and a weight-decay regularization term over parameters $\mathbf{\Theta}$, \model\ applies the following multi-task training loss with hyperparameters $\lambda_*$:
\begin{align}
    \label{eq:optim}
    \mathcal{L}=\lambda_0 \mathcal{L}_{bpr} + \lambda_1 \mathcal{L}_{p-kd} + \lambda_2 \mathcal{L}_{e-kd} + \lambda_3 \mathcal{L}_{u-reg} + \lambda_4 \|\mathbf{\Theta}\|_\text{F}^2.
\end{align}

\section{Evaluation}
\label{sec:exp}
We conduct extensive experiments on our \model\ framework, aiming to answer the following research questions (RQs):
\begin{itemize}[leftmargin=*]
    \item \textbf{RQ1}: How is the performance of \model\ after the model pruning, compared to existing recommendation methods?
    \item \textbf{RQ2}: How efficient is our pruned GNN, compared to baselines?%
    \item \textbf{RQ3}: How do the components of the proposed \model\ impact the recommendation performance of the pruned GNN?%
    \item \textbf{RQ4}: How do the pruning ratios impact the recommendation performance and the efficiency of the pruned GNN?%
    \item \textbf{RQ5}: Can the proposed \model\ framework alleviate the over-smoothing effect with its hierarchical knowledge distillation?%
    \item \textbf{RQ6}: Can our \model\ effectively identify the redundant and noisy information in the user-item interaction graph?
\end{itemize}

\subsection{Experimental Settings}
\subsubsection{\bf Datasets}
\begin{table}
    \centering 
    \small
    \caption{Statistical details of experimental datasets.}
    \label{tab:stat}
    \vspace{-0.1in}
    \begin{tabular}{ccccc}
    \hline
    Dataset &\# Users  &\# Items  &\# Interactions &Interaction Density \\
    \hline
    \hline
    Gowalla &25557  &19747  &294983 & $5.85 \times 10^{-4}$ \\
    Yelp    &42712  &26822  &182357 &  $1.59 \times 10^{-4} $ \\
    Amazon  &76469  &83761  &966680  & $1.51 \times 10^{-4} $  \\
    \hline
    \end{tabular}  
    \vspace{-0.1in}
\end{table}

\model\ is evaluated using three real-world datasets: Gowalla, Yelp, and Amazon. The \textbf{Gowalla} dataset contains user check-in records at geographical locations from January to June 2010, obtained from the Gowalla platform. \textbf{Yelp} dataset is obtained from Yelp platform and contains user ratings on venues from January to June 2018. The \textbf{Amazon} dataset contains people's ratings of books on the Amazon platform, during 2013. Following~\cite{xia2023graph_less}, we filter out users and items with less than three interactions, and splitting the original datasets into training, validation, and test sets by 70:5:25. Additionally, we convert ratings into binary implicit feedback, following~\cite{he2020lightgcn}. The data statistics are listed in Table~\ref{tab:stat}.

\subsubsection{\bf Evaluation Protocols}
We follow common evaluation protocols for recommendation
~\cite{wang2019neural, zhang2022incorporating}. We rank all uninteracted items with the positive items from test set for each user, a method known as full-rank evaluation. We use two common metrics, \textit{Recall@N} and \textit{NDCG@N}~\cite{wang2023diffusion, wu2021self} with values of $N=20$ and $40$.%

\subsubsection{\bf Baselines}
We compare \model\ to 18 baselines from diverse categories, including factorization method (\textbf{BiasMF}~\cite{koren2009matrix}), deep neural CF methods (\textbf{NCF}~\cite{he2017neural}, \textbf{AutoR}~\cite{sedhain2015autorec}), graph-based methods (\textbf{GCMC}~\cite{berg2017graph}, \textbf{PinSage}~\cite{ying2018graph}, \textbf{STGCN}~\cite{zhang2019star}, \textbf{NGCF}~\cite{wang2019neural}, \textbf{GCCF}~\cite{chen2020revisiting}, \textbf{LightGCN}~\cite{he2020lightgcn}, \textbf{DGCF}~\cite{wang2020disentangled}), self-supervised recommenders (\textbf{SLRec}~\cite{yao2021self}, \textbf{SGL}~\cite{wu2021self}, \textbf{NCL}~\cite{lin2022improving}, \textbf{SimGCL}~\cite{yu2022graph}, \textbf{HCCF}~\cite{xia2022hypergraph}), and compressed CF approaches (\textbf{GLT}~\cite{chen2021unified}, \textbf{UnKD}~\cite{chen2023unbiased}, \textbf{SimRec}~\cite{xia2023graph_less}).

\begin{table*}[t]
    \centering
    \caption{Overall performance comparison on Gowalla, Yelp, and Amazon datasets in terms of \textit{Recall@N} and \textit{NDCG@N}}
    \label{tab:main_tab}
    \vspace{-0.1in}
    \footnotesize
    \setlength{\tabcolsep}{0.6mm}
    \begin{tabular}{|c|c|c|c|c|c|c|c|c|c|c|c|c|c|c|c|c|c|c|c|c|c|}
    \hline
    Data                                          & Metric    & BiasMF & NCF    & AutoR  & PinSage & STGCN  & GCMC   & NGCF   & GCCF   & LightGCN & DGCF   & SLRec  & NCL    & SGL    & HCCF   & SimGCL &GLT &UnKD   & SimRec   & Ours \\ \hline
    \multicolumn{1}{|c|}{\multirow{4}{*}{Amazon}} & Recall@20 & 0.0324 & 0.0367 & 0.0525 & 0.0486  & 0.0583 & 0.0837 & 0.0551 & 0.0772 & 0.0868   & 0.0617 & 0.0742 & 0.0955 & 0.0874 & 0.0885 & 0.0921 &0.0901 & 0.0947 &0.1067 &  \textbf{0.1189}     \\ \cline{2-21} 
    \multicolumn{1}{|c|}{}                        & NDCG@20   & 0.0211 & 0.0234 & 0.0318 & 0.0317  & 0.0377 & 0.0579 & 0.0353 & 0.0501 & 0.0571   & 0.0372 & 0.0480 & 0.0623 & 0.5690 & 0.0578 & 0.0605 &0.0585 & 0.0607 
      & 0.0734 &  \textbf{0.0820}     \\ \cline{2-21} 
    \multicolumn{1}{|c|}{}                        & Recall@40 & 0.0578 & 0.0600 & 0.0826 & 0.0773  & 0.0908 & 0.1196 & 0.0876 & 0.1175 & 0.1285   & 0.0912 & 0.1123 & 0.1409 & 0.1312 & 0.1335 &0.1367  &0.1355 & 0.1376 & 0.1535 &  \textbf{0.1677}      \\ \cline{2-21} 
    \multicolumn{1}{|c|}{}                        & NDCG@40   & 0.0293 & 0.0306 & 0.0415 & 0.0402  & 0.0478 & 0.0692 & 0.0454 & 0.0625 & 0.0697   & 0.0468 & 0.0598 & 0.0764 & 0.0704 & 0.0716 &0.0730  &0.0725 & 0.0745 & 0.0879 & \textbf{0.0969}     \\ \hline%
    \multirow{4}{*}{Gowalla}                      & Recall@20 & 0.0867 & 0.1019 & 0.1477 & 0.0985  & 0.1574 & 0.1863 & 0.1757 & 0.2012 & 0.2230   & 0.2055 & 0.2001 & 0.2283 & 0.2332 & 0.2293 & 0.2328 &0.2324 &0.2331 & 0.2434 & \textbf{0.2610}      \\ \cline{2-21} 
     & NDCG@20   & 0.0579 & 0.0674 & 0.0690 & 0.0809  & 0.1042 & 0.1151 & 0.1135 & 0.1282 & 0.1433   & 0.1312 & 0.1298 & 0.1478 & 0.1509 & 0.1482 & 0.1506 &0.1464 &0.1496 & 0.1592 &   \textbf{0.1684}  \\ \cline{2-21} 
     & Recall@40 & 0.1269 & 0.1563 & 0.2511 & 0.1882  & 0.2318 & 0.2627 & 0.2586 & 0.2903 & 0.3181   & 0.2929 & 0.2863 & 0.3232 & 0.3251 & 0.3258 & 0.3276 &0.3269 &0.3301 & 0.3399 &  \textbf{0.3597}     \\ \cline{2-21} 
     & NDCG@40   & 0.0695 & 0.0833 & 0.0985 & 0.0994  & 0.1252 & 0.1390 & 0.1367 & 0.1532 & 0.1670   & 0.1555 & 0.1540 & 0.1745 & 0.1780 & 0.1751 & 0.1772 &0.1730 &0.1766 & 0.1865 &     \textbf{0.1962}   \\ \hline%
    \multirow{4}{*}{Yelp}                         & Recall@20 & 0.0198 & 0.0304 & 0.0491 & 0.0510  & 0.0562 & 0.0584 & 0.0681 & 0.0742 & 0.0761   & 0.0700 & 0.0665 & 0.0806 & 0.0803 & 0.0789 & 0.0788 &0.0812 &0.0819 & 0.0823 &  \textbf{0.0879}      \\ \cline{2-21} 
     & NDCG@20   & 0.0094 & 0.0143 & 0.0222 & 0.0245  & 0.0282 & 0.0280 & 0.0336 & 0.0365 & 0.0373   & 0.0347 & 0.0327 & 0.0402 & 0.0398 & 0.0391 & 0.0395 &0.0400 &0.0392 & 0.0414 &   \textbf{0.0443}   \\ \cline{2-21} 
     & Recall@40 & 0.0307 & 0.0487 & 0.0692 & 0.0743  & 0.0856 & 0.0891 & 0.1019 & 0.1151 & 0.1175   & 0.1072 & 0.1032 & 0.1230 & 0.1226 & 0.1210 & 0.1213 &0.1249 &0.1202 & 0.1251 & \textbf{0.1328}     \\ \cline{2-21} 
     & NDCG@40   & 0.0120 & 0.0187 & 0.0268 & 0.0315  & 0.0355 & 0.0360 & 0.0419 & 0.0466 & 0.0474   & 0.0437 & 0.0418 & 0.0505 & 0.0502 & 0.0492 & 0.0498 &0.0507 &0.0493 & 0.0519 & \textbf{0.0553}    \\ \hline
    \end{tabular}%
    \vspace{-0.1in}
\end{table*}

\subsubsection{\bf Hyperparameter Settings}
We implement \model\ with PyTorch, using Adam optimizer and Xavier initializer with default parameters. For all models, the training batch size is set to 4096 and the embedding size is 32 by default. For all GNN-based models, we set the layer number to 2. Weights $\lambda_0, \lambda_1, \lambda_2$ in \model are tuned from $\{1e^{-k}|k=0,1,...,4\}$. And $\lambda_3$ is tuned in a wider range which additionally contains $\{1e^{-5}, 1e^{-6}\}$. The weight $\lambda_4$ for weight-decay regularization is selected from $\{1e^{-k}|k=3,4,...,9\}$. All temperature coefficients are chosen from $\{1e^{-k}, 3e^{-k}, 5e^{-k}|k=-1,0,1,2\}$. Baseline methods are implemented using their released code with grid search for hyperparameter tuning. The efficiency test is conducted on a device with an NVIDIA GeForce RTX 3090 GPU.

\subsection{Performance Comparison (RQ1)}
We first compare \model\ to baselines on recommendation accuracy. The results are in Table~\ref{tab:main_tab}. We make the following observations:
\begin{itemize}[leftmargin=*]
    \item \textbf{Superior performance of \model}: The proposed model \model\ surpasses all baselines across different categories, including simple neural CF, graph-based recommenders, self-supervised methods, and compression methods. This superiority in performance demonstrates that our learnable pruning framework and hierarchical distillation paradigm not only maintain prediction accuracy after model compression but also enhance existing recommendation frameworks. The effective elimination of noise and redundancy in the interaction graph and embedding parameters contributes to these performance improvements.\\\vspace{-0.12in}
    \item \textbf{Drawbacks of CF without model compression}: When comparing the best-performing CF methods, such as self-supervised CF techniques like SGL, HCCF, and SimGCL, to compression methods like UnKD and SimRec, it is evident that CF methods without model compression fall short in terms of recommendation accuracy. This discrepancy can be attributed to the debiasing and anti-over-smoothing effects embedded in the knowledge distillation process of UnKD and SimRec. This suggests that model compression techniques, such as knowledge distillation, can go beyond improving model efficiency. They can also address adverse factors present in observed data and modeling frameworks, such as data bias, noise, and over-smoothing effects.\\\vspace{-0.12in}
    \item \textbf{Importance of explicit noise elimination}: While UnKD and SimRec refine the distilled model by addressing bias and over-smoothing effects in GNN-based CF, they rely solely on high-level supervision methods. In contrast, our \model\ explicitly identifies and eliminates fine-grained noisy and redundant elements within the model, such as edges and embedding entries. This empowers our \model\ with notable strength in recommender refinement, leading to significant performance superiority.
\end{itemize}

\subsection{Efficiency Test (RQ2)}
To assess the model efficiency, we evaluate the memory and computational costs of \model\ and baselines.
The compared baselines include NGCF, GCCF, HCCF, and existing GNN compression method UnKD. Our \model\ is tested with different preservation ratios.
In Figure~\ref{fig:efficiency}, the results are presented relative to the performance of NGCF. We deduce the following observations:
\begin{figure}[t]
    \centering
    \subfigure[Storage costs.]{ \includegraphics[width=0.47\columnwidth]{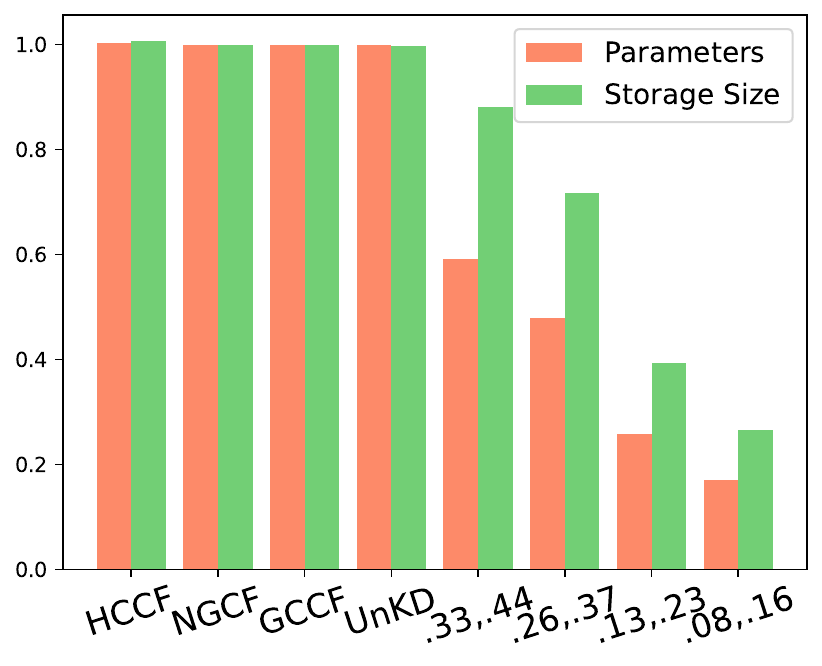}\label{fig:params_size}}
    \subfigure[Time costs.]{\includegraphics[width=0.47\columnwidth]{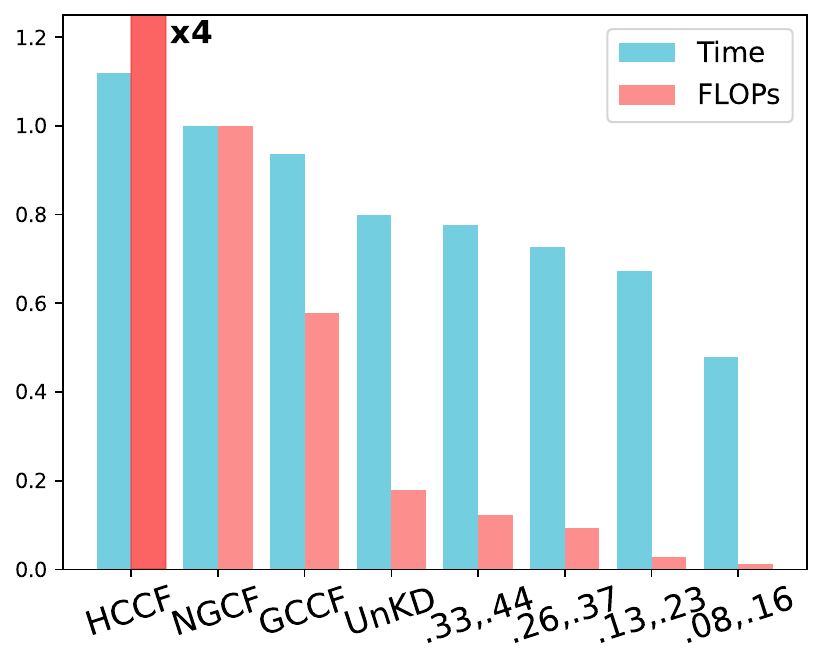}}
    \vspace{-0.18in}
    \caption{Disk storage and time costs of baselines and our \model\ under different preservation ratios (\eg~.33, .44 denote preserving 33\% embedding entries and 44\% edges).}
    \label{fig:efficiency}
    \vspace{-0.15in}
\end{figure}
\begin{itemize}[leftmargin=*]
    \item \textbf{Simplified GNNs}. Despite simplifying the GNN architecture by removing transformations and activations, some GNN methods like GCCF fail to significantly reduce memory and time costs related to graph storage and information propagation. Consequently, the costs of GCCF remain comparable to those of NGCF. This demonstrates the limitation of architectural simplifications in improving efficiency for graph-based recommendation.\\\vspace{-0.12in}
    \item \textbf{SSL-enhanced GNNs}. SSL techniques have been utilized to enhance graph recommenders by generating self-supervision signals. However, it is important to note that these methods may introduce additional operations, leading to increased memory and time costs. This is exemplified by the performance of HCCF, where utilizing extra hypergraph propagation necessitates more FLOPs and yields a noticeable increase in computational time.\\\vspace{-0.12in}
    \item \textbf{Existing compressed GNNs}. UnKD has been successful in achieving efficiency improvements, particularly in terms of computational time. However, when comparing UnKD to \model, a significant disadvantage becomes evident. This limitation arises from UnKD's lack of explicit identification and removal of redundancy and noise in the GNN model. As a result, UnKD is unable to prune a larger portion of the GNN to achieve superior efficiency improvements like our \model\ framework does.\\\vspace{-0.12in}
    \item \textbf{Efficiency of \model}. The results demonstrate a significant memory reduction of 70\% in \model, considering both the parameter number and storage size. Moreover, there is an impressive reduction of over 90\% in FLOPs during forward propagation and an over 50\% reduction in physical prediction time. These efficiency optimizations can be attributed to two key aspects. \textbf{Firstly}, the learnable GNN pruning paradigm accurately removes redundant and noisy information from the GNN. This facilitates efficient utilization of computational resources. \textbf{Secondly}, our learnable pruning mechanism is supervised by the hierarchical KD, which incorporates multi-dimensional alignment and high-order structure augmentation. This maximizes the retention of performance, allowing for more extensive pruning of parameters.
\end{itemize}

\subsection{Ablation Study (RQ3)}
\begin{table}[t]
    \centering
    \caption{Ablation study of \model\ measured by Recall@20.}
    \label{tab:ablation_study}
    \small
    \setlength{\tabcolsep}{0.6mm}
    \vspace{-0.1in}
    \begin{tabular}{c|l|c|c|c|c|c|c|c|c}
        \hline
        \multicolumn{2}{c|}{Dataset} & \multicolumn{4}{c|}{Gowalla} & \multicolumn{4}{c}{Yelp} \\ \hline
        \multicolumn{2}{c|}{Ratio $\mate/\sete$} & \multicolumn{1}{c|}{.33/.44} & \multicolumn{1}{c|}{.26/.37} & \multicolumn{1}{c|}{.11/.19} & .08/.16 & \multicolumn{1}{c|}{.33/.77} & \multicolumn{1}{c|}{.26/.74} & \multicolumn{1}{c|}{.11/.60} & .08/.57 \\ \hline \hline

        \multirow{4}{*}{Prn} & \textasciitilde EmbP           & \multicolumn{1}{c|}{0.2197}   & \multicolumn{1}{c|}{0.1946}   & \multicolumn{1}{c|}{0.0741}   & 0.0586   & \multicolumn{1}{c|}{0.0809}   & \multicolumn{1}{c|}{0.0737}   & \multicolumn{1}{c|}{0.0434}   & 0.0357  \\ 
        \cline{2-10}
        &\textasciitilde EdgeP         & \multicolumn{1}{c|}{0.2418}   & \multicolumn{1}{c|}{0.2255}   & \multicolumn{1}{c|}{0.1341}   & 0.1133  & \multicolumn{1}{c|}{0.0867}   & \multicolumn{1}{c|}{0.0850}   & \multicolumn{1}{c|}{0.0745}    & 0.0709  \\ 
        \cline{2-10}
        & \textasciitilde BothP          & \multicolumn{1}{c|}{0.1800}   & \multicolumn{1}{c|}{0.1434}   & \multicolumn{1}{c|}{0.0556}   & 0.0421  & \multicolumn{1}{c|}{0.0736}   & \multicolumn{1}{c|}{0.0661}   & \multicolumn{1}{c|}{0.0311}   & 0.0231  \\
        \cline{2-10}
        & BnEdge & \multicolumn{1}{c|}{0.2210}   & \multicolumn{1}{c|}{0.2021}   & \multicolumn{1}{c|}{0.1280}   & 0.1165   & \multicolumn{1}{c|}{0.0872}   & \multicolumn{1}{c|}{0.0858}   & \multicolumn{1}{c|}{0.0775}   & 0.0754  \\ \hline
        
        \multirow{3}{*}{KD} & -BiAln & \multicolumn{1}{c|}{0.2350}   & \multicolumn{1}{c|}{0.2281}   & \multicolumn{1}{c|}{0.1934}   & 0.1812  & \multicolumn{1}{c|}{0.0810}   & \multicolumn{1}{c|}{0.0801}   & \multicolumn{1}{c|}{0.0666}   & 0.0650  \\
        \cline{2-10}
        & -IntKD & \multicolumn{1}{c|}{0.2607}   & \multicolumn{1}{c|}{0.2571}   & \multicolumn{1}{c|}{0.2100}   & 0.1890  & \multicolumn{1}{c|}{0.0822}   & \multicolumn{1}{c|}{0.0825}   & \multicolumn{1}{c|}{0.0788}   & 0.0769  \\ 
        \cline{2-10}
        & -ImpD & \multicolumn{1}{c|}{0.2593}   & \multicolumn{1}{c|}{0.2564}   & \multicolumn{1}{c|}{0.2135}   & 0.1923  & \multicolumn{1}{c|}{0.0862}   & \multicolumn{1}{c|}{0.0861}   & \multicolumn{1}{c|}{0.0808}   & 0.0797  \\
        
        \hline\hline
        \multicolumn{2}{c|}{\model} & \multicolumn{1}{c|}{0.2610}   & \multicolumn{1}{c|}{0.2578}   & \multicolumn{1}{c|}{0.2162}   & 0.1966  & \multicolumn{1}{c|}{0.0879}   & \multicolumn{1}{c|}{0.0877}   & \multicolumn{1}{c|}{0.0856}   & 0.0842  \\ \hline
        \end{tabular}
    \vspace{-0.15in}
\end{table}

We investigate the effectiveness of \model's technical designs using Gowalla and Yelp data, with different pruning ratios. The results are shown in Table~\ref{tab:ablation_study}. We make the following observations.

\noindent\textbf{Effectiveness of the GNN pruning techniques}.
\begin{itemize}[leftmargin=*]
    \item \textbf{\textasciitilde EmbP}, \textbf{\textasciitilde EdgeP}, \textbf{\textasciitilde BothP}: We replace the learnable pruning with random dropping. The three variants replace embedding pruning, edge pruning, and both, respectively. Significant performance drop can be observed under different pruning ratios, indicating the effectiveness of our learnable pruning in identifying the essential embedding entries and edges. Especially, when dropping with high ratios (\eg~preserving only 11\% and 8\% entries), the prediction ability of the random variants experiences a destructive (over 70\%) decay, while \model\ preserves most of its accuracy.\\\vspace{-0.12in}
    \item \textbf{BnEdge}: To study the effect of learned edge weights $\matw^s$, BnEdge uses binary edge weights instead of $\matw^s$ during GNN propagation. Though it maintains the learnable pruning process unchanged, a noticeable degradation can be observed. This suggests the crucial role of learned weights. They not only identify which edges to prune, but also effectively preserve the pruned information.
\end{itemize}

\noindent\textbf{Effectiveness of knowledge distillation}.
\begin{itemize}[leftmargin=*]
    \item \textbf{-BiAln}: To assess the significance of the KD constraints for effective pruning, we remove the bilevel alignment, including the prediction-level and embedding-level KD. The notable performance drop verifies the importance of aligning the teacher model with the pruned model, to effectively retain model performance.\\\vspace{-0.12in}
    \item \textbf{-IntKD}: This variant removes the intermediate KD layer in \model. As a result, its performance notably deteriorates, particularly on the Yelp dataset. The increased importance of this module for Yelp can be attributed to the higher sparsity of the dataset. In such cases, the intermediate KD layer is able to seek more edges from high-order relations to enrich the small edge set.\\\vspace{-0.12in}%
    \item \textbf{-ImpD}: This variant removes the importance distillation, and the results confirm the benefits of incorporating learned edge weights and predictions from the intermediate KD layer model into the decision-making process of edge dropping.    
\end{itemize}
\subsection{Influence of Pruning Ratios (RQ4)}
\begin{figure}
    \centering
    \subfigure{\includegraphics[width=\columnwidth]{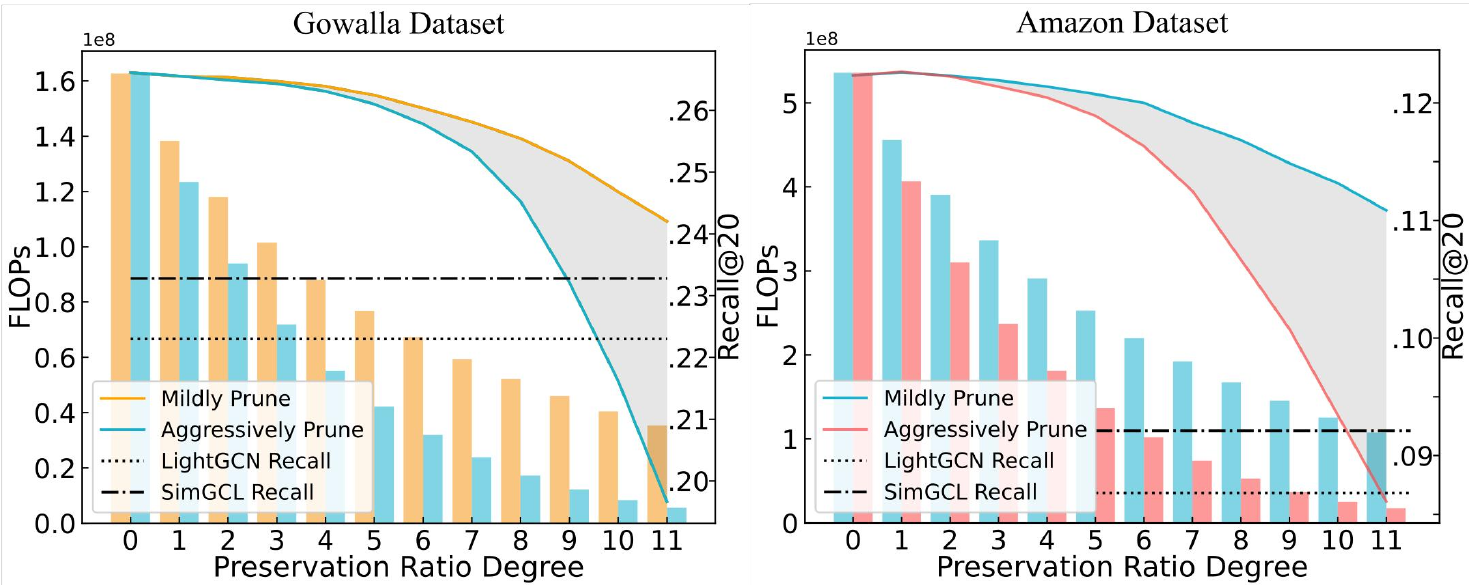}}\vspace{-0.2in}\vspace{0.1in}
    \subfigure{
        \includegraphics[width=0.99\columnwidth]{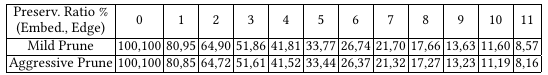}
    }\vspace{-0.12in}    
    \caption{Computational FLOPs (bars) and recommendation performance (lines) \wrt~different preservation ratios.}
    \label{fig:hyper_params}
    \vspace{-0.15in}
\end{figure}
In this experiment, we investigate the impact of pruning ratios for edges and embedding entries on both model performance and efficiency. Figure~\ref{fig:hyper_params} shows the evaluated model performance and computing FLOPs (floating point operations) during forward propagation across various preservation ratios. We present two pruning schemes: a mild pruning scheme that removes fewer graph edges in the GNN, and an aggressive pruning scheme that removes more edges. Based on the results, we draw the following observations:
\begin{itemize}[leftmargin=*]
    \item \textbf{Performance change}. As we discard a larger number of embedding entries and graph edges, we observe a continuous decline in performance. However, it is noteworthy that even when a substantial portion of the GNN model is removed, our \model\ consistently maintains a high level of recommendation performance compared to SimGCL and LightGCN. This resilience can be attributed to the hierarchical KD, which effectively aligns the predictions of the student model with those of the well-performing teacher model through bilevel alignment, and the importance distillation that gives the optimal dropping strategies. Additionally, the intermediate KD layer with structure augmentation further enhances the recommendation ability by incorporating more edges sampled from high-order relations. These features collectively contribute to the robust performance of \model.\\\vspace{-0.12in}
    \item \textbf{Efficiency change}. As the pruning ratio increases, \model\ exhibits a significant decrease in FLOPs. This confirms the effectiveness of enhancing GNN efficiency by pruning embeddings and structures. Specifically, our \model\ achieves a FLOPs reduction of 90\% during forward propagation while maintaining comparable performance to SimGCL, and a FLOPs reduction of 95\% while performing similarly to LightGCN. These substantial reductions in FLOPs highlight the effectiveness of our learnable pruning strategy in minimizing computational operations.\\\vspace{-0.12in}
    \item \textbf{Differences across datasets}. Furthermore, it is worth mentioning that our \model\ demonstrates better preservation of recommendation performance when pruning the same proportion of information on the Amazon dataset compared to the Gowalla dataset. This observation suggests the presence of more redundancy or noise in the Amazon data, which aligns with the larger number of edges and users/items present in the Amazon dataset.
\end{itemize}

\subsection{Anti-Over-Smoothing Effect Study (RQ5)}

\begin{figure}[t]
\centering
    \subfigure[LightGCN]{ 
        \includegraphics[width=0.244\columnwidth]{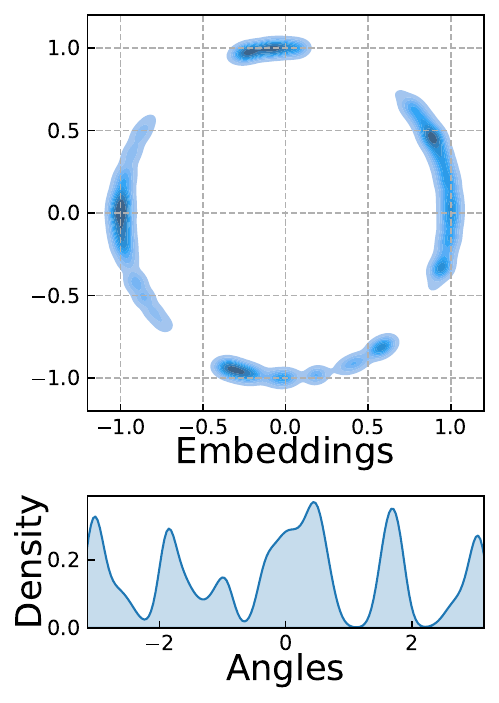}
        \label{fig:lightgcn_dis}
    }
    \subfigure[SGL]{
        \hspace{-0.09in}\includegraphics[width=0.244\columnwidth]{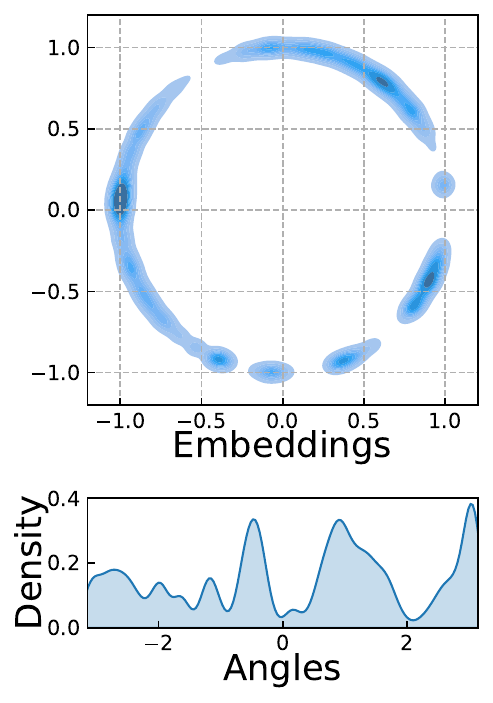}
        \label{fig:sgl_dis}
    }
    \subfigure[SimGCL]{
        \hspace{-0.09in}\includegraphics[width=0.244\columnwidth]{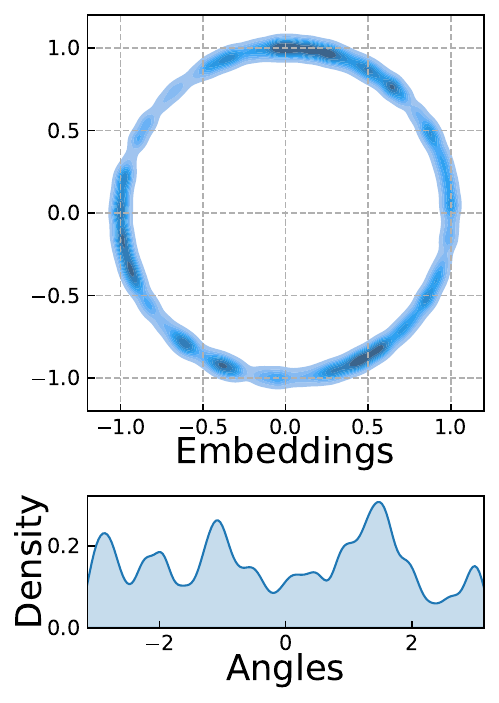}
        \label{fig:simgcl_dis}
    }
    \subfigure[Ours]{
        \hspace{-0.09in}\includegraphics[width=0.244\columnwidth]{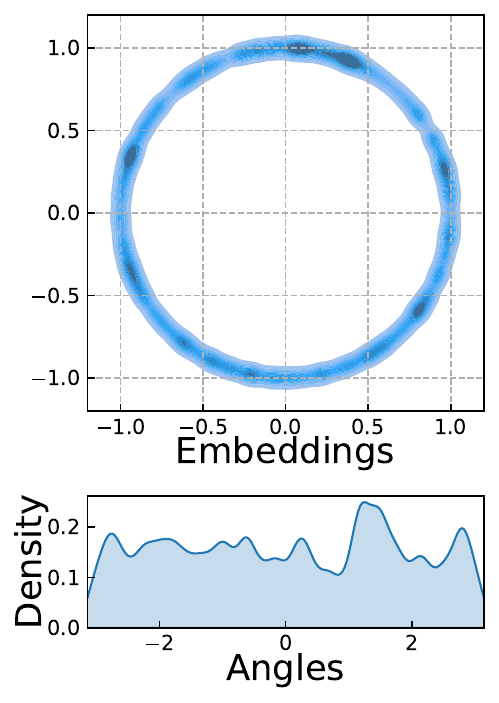}
        \label{fig:our_dis}
    }
    \vspace{-0.15in}
    \caption{Embedding distributions in the 2-D space and in the 1-D angle space for Yelp dataset, estimated by KDE.}
    \label{fig:embed_dis}
    \vspace{-0.1in}
\end{figure}

\begin{table}[t]
    \centering
    \caption{MAD among popular nodes from Yelp and Gowalla.}
    \label{tab:mad}
    \vspace{-0.1in}
    \resizebox{\columnwidth}{!}{
    \begin{tabular}{c|c|c|c|c|c|c|c}
        \hline

        \hline
        Datasets & GCCF   & LightGCN & NCL & SGL    & SimGCL & SimRec & Ours            \\ \hline \hline
        Yelp     & 0.8747  & 0.8819 & 0.8929   & 0.8643 & 0.9103 & 0.9272 & \textbf{0.9404} \\ \hline
        Gowalla  & 0.8206 & 0.8269 & 0.8236    & 0.7760 & 0.8595 & 0.8406 & \textbf{0.8742} \\ \hline

        \hline
    \end{tabular}
    }
    \vspace{-0.15in}
\end{table}

To assess the ability of \model\ to mitigate the over-smoothing effect of GNNs during the pruning process, we compare the distribution uniformity of our model's embeddings with those of baseline methods. This comparison is conducted in two dimensions.
\begin{itemize}[leftmargin=*]
    \item \textbf{Visualization of Embedding Distribution}. 
    From the embedding distributions plotted in Figure~\ref{fig:embed_dis}, we can observe that: 
    \textbf{i)} The clustering effect observed in both 2-D plots and angle plots is notably stronger for LightGCN, demonstrating the severe over-smoothing effect resulting from the iterative embedding smoothing paradigm. 
    \textbf{ii)} To address this issue, SGL and SimGCL incorporate contrastive learning to enhance the distribution uniformity of embeddings. Both methods exhibit higher uniformity in the estimated distribution compared to LightGCN, with SimGCL exhibiting some superiority due to its less-random augmentation design.
    \textbf{iii)} Compared to SimGCL, our \model\ exhibits even fewer dark regions in the embedding distribution ring, indicating higher uniformity. This advantage becomes more apparent in the angle-based plot, where the low probabilities are much closer to the high ones in \model. This observation strongly indicates a higher anti-over-smoothing ability of our \model, 
    which can be ascribed to the sparsification effect caused by embedding pruning, and the uniformity constraint in our \model.
    \\\vspace{-0.12in}
    \item \textbf{Mean Average Distance (MAD) Values}.
    We further evaluate the MAD values~\cite{chen2020measuring, xia2023graph_less} in Table~\ref{tab:mad}, from which we draw the following observations: 
    \textbf{i)} The GNN-based CF paradigms GCCF and LightGCN generally exhibit lower MAD values compared to other methods that employ contrastive learning. This observation highlights the inherent over-smoothing issue in propagation-based graph neural encoders.
    \textbf{ii)} For the other baselines, we observe that NCL and SGL show lower MAD values, indicating a stronger over-smoothing effect. This sheds light on the limitations of their random structure augmentation methods, which are susceptible to the influence of data noise.
    \textbf{iii)} The superiority of SimGCL and SimRec validates their effective design of pushing all embeddings apart. In comparison, our \model\ achieves further advancements by constructing meaningful positive sample pairs using node-wise similarity in embedding pruning. This technique effectively enhances positive relation learning in a learnable manner.
\end{itemize}

\subsection{Noise and Redundancy Identification (RQ6)}

\begin{figure}
    \centering
    \includegraphics[width=\columnwidth]{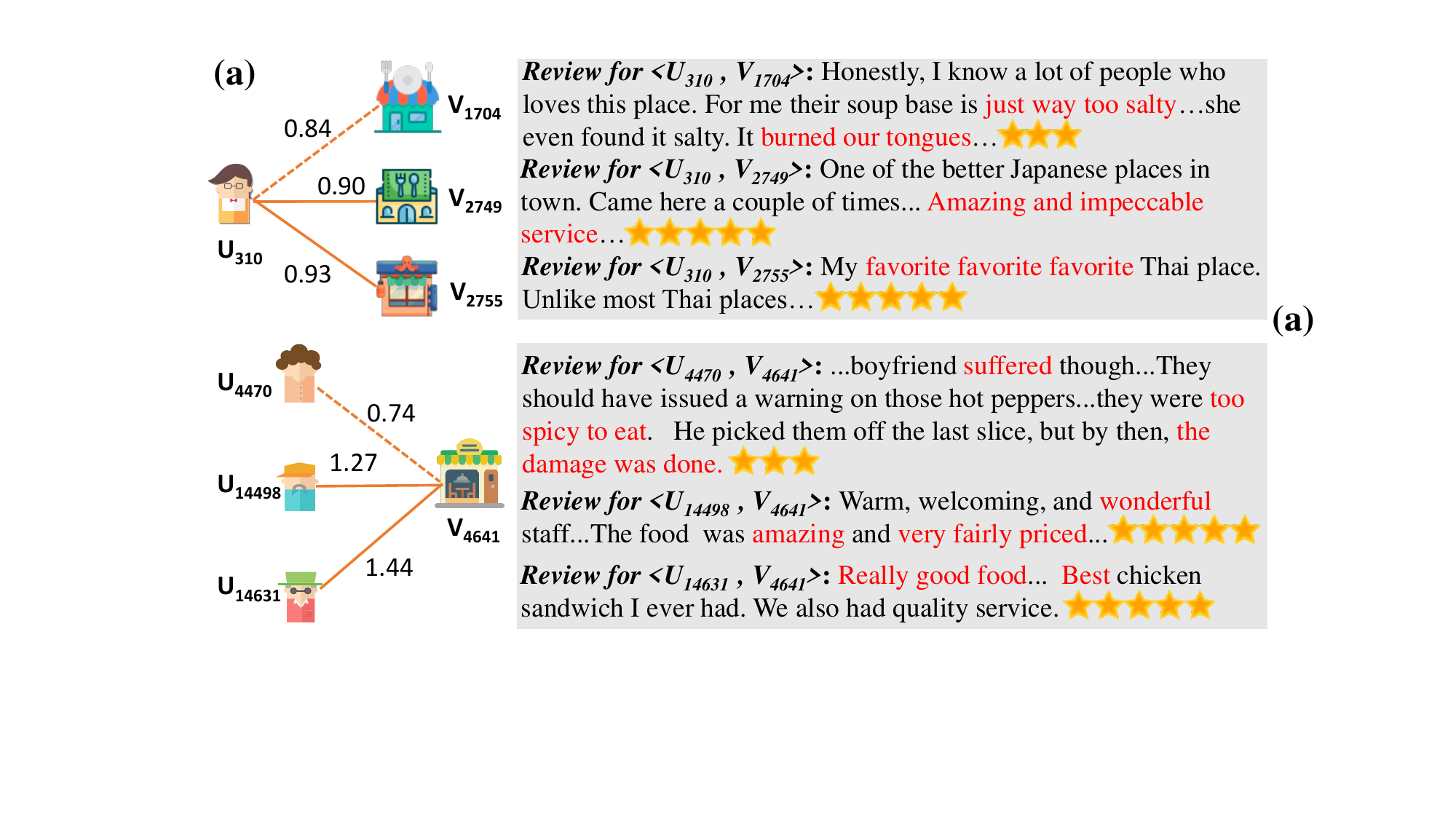}
    
    \includegraphics[width=\columnwidth]{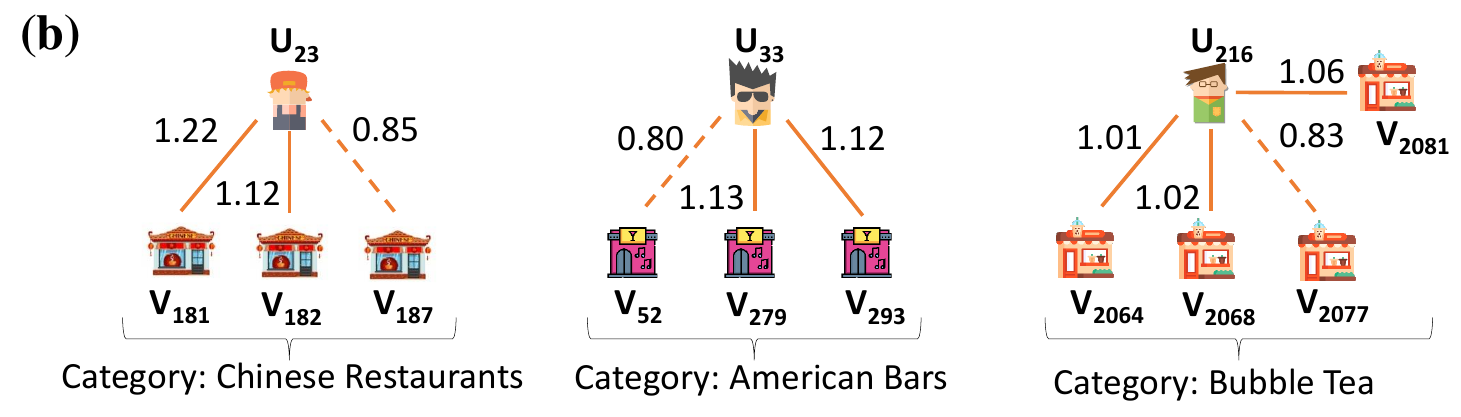}
    \vspace{-0.25in}
    \caption{Investigation on the capability of (a) noise pruning and (b) redundancy pruning for our \model\ framework.}
    \label{fig:case_study}
    \vspace{-0.15in}
\end{figure}

We explore \model's capacity to trim noise and redundancy in interaction data. The results are detailed in Figure~\ref{fig:case_study}.%

\noindent\textbf{Noise Pruning}. In Figure~\ref{fig:case_study}(a), two sets of decision weights in $\bar{\matw}^s$ for left-side edges are depicted alongside users' text reviews and ratings for corresponding items on the right. Notably, these reviews and ratings were not exposed to our \model. Our results show that \model\ assigns low weights to interactions like $<u_{310}~,~v_{1704}>$ and $<u_{4470}~,~v_{4641}>$, aligning with users' negative feedback (e.g., "too salty."). In the context of graph CF, such negative feedback instances are viewed as regular user-item interactions, possibly adversely affecting user preference modeling. Frequent similar observations in our results show that \model\ effectively identifies and addresses noise in the graph structure, thereby improving the pruning effect of GNN-based recommendation.\\\vspace{-0.12in}

\noindent\textbf{Redundancy Pruning}. 
In Figure~\ref{fig:case_study}(b), some representative cases demonstrate the efficacy of redundancy pruning in \model, where three users interact with multiple venues sharing the same categories like Chinese restaurants and American bars, reflecting redundant user interest information. Despite being category-agnostic, \model\ identifies these similarities, assigning lower weights to some of the redundant items. This encourages the pruning algorithm to eliminate the redundancy, thereby enhancing model efficiency. Moreover, thanks to the learnable edge weights in the intermediate KD layer, \model\ preserves preference strength for each interest, rather than relying on item counts of each interest.

\section{Related Work}
\label{sec:relate}

\subsection{Graph Neural Recommender Systems}
Graph neural networks (GNNs) have emerged as foundational architectures for recommendation systems. Early works such as NGCF~\cite{wang2019neural} and GCMC~\cite{berg2017graph} introduced graph convolutional networks (GCNs) for collaborative recommendation. Subsequent studies include STGCN~\cite{zhang2019star}, which integrates an autoencoding architecture within the GNN encoder, and DGCF~\cite{wang2020disentangled}, which incorporates a representation disentanglement module into graph-based collaborative filtering. LightGCN~\cite{he2020lightgcn} and GCCF~\cite{chen2020revisiting} emphasize the redundancy in prior graph neural architectures and achieve improved performance by eliminating both non-linear and linear mappings.

Recently, self-supervised learning (SSL) has gained attention for its ability to generate rich supervision signals and address the data sparsity problem in recommendation. Contrastive learning (CL)-based graph CF (\eg~SGL~\cite{wu2021self}, SimGCL~\cite{yu2022graph}, DirectAU~\cite{wang2022towards}, AdaGCL~\cite{jiang2023adaptive}) is a popular SSL technique that effectively learns a uniform distribution to counter the over-smoothing effect of GNNs. HCCF~\cite{xia2022hypergraph} and NCL~\cite{lin2022improving} introduce additional encoding views to enrich graph CL. In addition, graph-based recommendation has also been enhanced with generative SSL techniques based on masked autoencoding, such as AutoCF~\cite{xia2023automated} and DGMAE~\cite{ren2023distillation}.

Despite the substantial enhancements in recommendation performance due to GNN advancements, an inherent limitation remains in the inefficiency of GNN's extensive information propagation and node-specific parameters. In this context, our \model\ aims to effectively prune redundant and noisy components of GNNs while preserving high performance through distillation constraints.

\subsection{Model Compression for Graph Models}
To enhance the scalability of GNNs, prior works have utilized random node and edge sampling techniques for large graphs (e.g., PinSAGE~\cite{ying2018graph}, HGT~\cite{hu2020heterogeneous}). However, these random strategies do not ensure the preservation of crucial information and may significantly affect model performance. In response, several approaches have emerged to better retain important patterns from the original model. GLT~\cite{chen2021unified} advocates for preserving only the essential edges by learning their importance to downstream task performance. Other studies improve compression supervision through knowledge distillation. GLNN~\cite{zhang2021graph} and SimRec~\cite{xia2023graph_less} propose distilling efficient student models based on MLP from heavier GNNs. UnKD~\cite{chen2023unbiased} further mitigates bias in the KD process using a stratified distillation strategy. Additionally, KD has been applied to compress recommenders based on non-GNN architectures (e.g.,~\cite{tang2018ranking, xia2022device}).

In contrast to previous approaches that broadly reduce model complexity by substituting GNNs with simpler architectures, our \model\ preserves robust topology extraction capabilities of GNNs. It achieves efficiency by explicitly identifying and eliminating redundancy and noise within GNN structures and embeddings. This strategy effectively mitigates misinformation in the graph while enhancing interpretability through pruned information.

\section{Conclusion}
\label{sec:conclusoin}
This paper introduces a novel pruning framework, \model, aimed at addressing scalability and robustness challenges in GNN-based collaborative filtering. \model\ explicitly models the probabilities of redundancy and noise for each edge and embedding parameter within the GNN recommender, enabling precise pruning of misinformation. It is driven by innovative hierarchical distillation objectives that leverage high-order relations and multi-level distillation to enhance performance retention. Extensive experiments demonstrate that \model\ outperforms baselines in recommendation performance, compression efficiency, and robustness.

\clearpage
\bibliographystyle{abbrv}
\balance
\bibliography{refs}
\clearpage
\appendix \section{Ethical Considerations}
\label{sec:appendix}

\subsection{Ethical Implications}
The proposed research on \model, a distillation-based GNN pruning framework, introduces innovative techniques for model compression in Graph Neural Networks (GNNs) to reduce model complexity
while preserving recommendation accuracy. While the advancements in this field are promising, there are ethical implications that need to be considered since graph-based recommendation systems often rely on sensitive user interaction data.
\begin{itemize}[leftmargin=*]
    \item \textbf{Privacy Considerations}. 
    GNN's utilization of user interaction data raises concerns about privacy. The pruning process must safeguard against unauthorized access to sensitive user information contained within the graph data. Besides, the removal of edges and embedding entries during compression should be conducted in a manner that does not inadvertently expose or retain identifiable user information.\\\vspace{-0.1in}
    \item \textbf{Security and Safety}.
    Pruning components based on learnable algorithms may introduce vulnerabilities that could compromise the integrity of the recommendation system, potentially leading to data breaches or manipulation. Moreover, aggressive pruning to achieve high compression rates might compromise the robustness of the GNN model, making it more susceptible to adversarial attacks or unexpected behaviors.
\end{itemize}

\subsection{Mitigation Strategies}
Below, we introduce some possible mitigation strategies.
\begin{itemize}[leftmargin=*]
    \item \textbf{Privacy-Preserving Techniques}. 
    Implement encryption and anonymization methods to protect user data while ensuring that the pruning process does not compromise individual privacy.\\\vspace{-0.1in}
    \item \textbf{Security Audits}.
    Conduct thorough security assessments to identify and address potential vulnerabilities introduced by the pruning framework, ensuring data integrity and system security.
    \item \textbf{Transparency and Accountability}.
    Maintain transparency in the pruning process, providing clear explanations of how components are pruned and enabling users to understand and challenge the recommendations made by the system.    
\end{itemize}
In conclusion, while \model\ shows promise in reducing model complexity while retaining recommendation performance, it is important for researchers and developers to prioritize ethical considerations to mitigate potential negative societal impacts and uphold the integrity and fairness of AI systems in recommendation.

\end{document}